\documentclass[aps,prb,reprint,groupedaddress,showpacs]{revtex4-1}

\usepackage{amsmath}
\usepackage{bm}
\usepackage{graphicx}

\newcommand{\ave}[1]{\left\langle #1 \right\rangle}
\newcommand{\bra}[1]{\left\langle #1 \right|}
\newcommand{\ket}[1]{\left| #1 \right\rangle}
\newcommand{\braket}[2]{\left\langle #1 \right| \left.\!\! #2 \right\rangle}
\newcommand{\cf}{\mathcal{F}}
\newcommand{\tf}{\widetilde{\mathcal{F}}}

\makeatletter
\AtBeginDocument{\@ifpackageloaded{natbib}{\ifNAT@numbers\if@filesw\immediate\write\@auxout{\string\global\string\NAT@numberstrue}\fi\fi}{}}
\makeatother
\begin{document}


\title{Odd-frequency pairing effect on the superfluid density and the Pauli
 spin susceptibility in spatially nonuniform spin-singlet superconductors}


\author{S. Higashitani}
\affiliation{Graduate School of Integrated Arts and Sciences,
 Hiroshima University,
 Kagamiyama 1-7-1, Higashi-Hiroshima 739-8521, Japan}


\date{\today}

\begin{abstract}
 A theoretical study is presented on the odd-frequency spin-singlet pairing
 that arises in nonuniform even-frequency superconductors as a consequence
 of broken translation symmetry. The effect of the odd-frequency pairing on
 the superfluid density and the spin susceptibility is analyzed by using the
 quasiclassical theory of superconductivity. It is shown that (1) the
 superfluid density is reduced by the formation of the odd-frequency singlet
 pairs and (2) the odd-frequency pairing increases the spin susceptibility
 even though its spin symmetry is singlet. The two unusual phenomena are
 related to each other through a generalized Yosida formula by taking into
 account both the even- and odd-frequency pairing effects.
\end{abstract}


\maketitle

\section{Introduction}

The concept of odd-frequency pairing offers  interesting symmetry aspects
of nonuniform superconductivity and superfluidity.\cite{TanakaJPSJ2012}
Although the odd-frequency pairing state was originally proposed as a
uniform superfluid state in bulk,\cite{BerezinskiiJETPL1974} it may also
emerge in, e.g., superconducting proximity structures. Bergeret, Volkov, and
Efetov pointed out, in their theoretical work on a
ferromagnet-superconductor proximity structure, that triplet $s$-wave pairs
are created in a ferromagnet attached to a conventional singlet $s$-wave
superconductor.\cite{BergeretPRL2001} In the ferromagnet, spin-rotation
symmetry is broken and the resulting singlet-triplet spin mixing generates
the triplet pairs from the singlet pairs penetrating from the
superconductor.  The Pauli principle requires that the triplet $s$-wave pair
amplitude be an odd function of the Matsubara frequency, and thus this pairing
state belongs to the odd-frequency symmetry class.  Similar odd-frequency
pairing takes place even in a normal metal when a superconductor is in
contact with it through a spin-active
interface.\cite{EschrigJPSJ2007,LinderPRL2009,LinderPRB2010} In proximity
structures, broken translation symmetry resulting from the presence of the
interface/surface provides another mechanism responsible for the emergence
of odd-frequency states. The symmetry breaking in real space causes mixing
of different orbital-parity states, so that  admixtures of
even- and odd-frequency states arise around the interface/surface.
\cite{TanakaPRL2007,TanakaPRB2007,HigashitaniJLTP2009,HigashitaniPRB2012}
This creation mechanism works without any magnetism and suggests a ubiquitous
existence of odd-frequency pairing states in nonuniform systems.

Recently, Yokoyama, Tanaka, and Nagaosa examined the effect of odd-frequency pairing
on the magnetic response of a normal metal--superconductor junction with
a spin-active interface.\cite{YokoyamaPRL2011} On the basis of
Usadel's dirty-limit theory,\cite{UsadelPRL1970,AlexanderPRB1985} it was shown that the
proximity-induced odd-frequency pairing state exhibits {\em paramagnetic}
Meissner response and gives rise to oscillation of the penetrating magnetic
field. The origin of this anomalous phenomenon can be found in the
dirty-limit formula for the superfluid fraction (the ratio of the superfluid
density $n_s$ to the total number density
$n$),\cite{UsadelPRL1970,AlexanderPRB1985}
\begin{align}
 \frac{n_s}{n} = \frac{2\tau_{\rm tr}}{\hbar}\frac{\pi}{\beta}\sum_{\epsilon_n}
 \left(
 -\frac{1}{2}{\rm Tr}\left[F(\epsilon_n)F(\epsilon_n)^*\right]
 \right),
 \label{Usadel-J-formula}
\end{align}
where $\tau_{\rm tr}$ is the transport mean free time, $\beta = 1/k_BT$ is
the inverse temperature, $\epsilon_n = (2n+1)\pi/\beta$ is the Matsubara frequency,
and $F(\epsilon_n)$ is an $s$-wave pair amplitude defined as a spin-space
matrix. Conventional $s$-wave superconductivity is described by $F_{\rm
singlet}(\epsilon_n) = f(\epsilon_n)i\sigma_2$ with $f(\epsilon_n)$ being an
even-frequency amplitude and $\sigma_2$ being the second component of the Pauli matrix
$\bm{\sigma} = (\sigma_1, \sigma_2, \sigma_3)$. The expression in
parentheses in Eq.\ \eqref{Usadel-J-formula} then gives the pair density
$|f(\epsilon_n)|^2$. In contrast, odd-frequency $s$-wave pairing is
characterized by $F_{\rm triplet}(\epsilon_n) =
\bm{f}(\epsilon_n)\cdot\bm{\sigma}i\sigma_2$. We then obtain the {\em
negative} pair density $-\bm{f}(\epsilon_n)\cdot\bm{f}(\epsilon_n)^*$ from
the same expression as above. This means that the odd-frequency pairs carry
paramagnetic Meissner current.  The negative pair density causes not only
the paramagnetic Meissner effect but also an unusual behavior of surface
impedance.\cite{AsanoPRL2011,AsanoPRB2012}

An anomaly resulting from  odd-frequency pairing also manifests itself  in Pauli spin
susceptibility $\chi$.\cite{HigashitaniPRL2013} It was predicted that
odd-frequency ($\uparrow\downarrow$$+$$\downarrow\uparrow$)-triplet pairing
in  nonuniform superfluid $^3$He {\em increases} the susceptibility $\chi$,
contrary to the conventional wisdom that antiparallel spin pairing reduces
$\chi$ in superfluids and superconductors. The question then naturally
arises and still remains whether the odd-frequency {\em singlet} pairing
also increases the susceptibility $\chi$. In bulk singlet $s$-wave
superconductors, the susceptibility $\chi^{\rm bulk}$ can be represented in
terms of the superfluid density $n_s^{\rm bulk}$ as
\begin{align}
 \frac{\chi^{\rm bulk}}{\chi_0^{}} = 1 - \frac{n_s^{\rm bulk}}{n}.
 \label{Yosida-formula}
\end{align}
This so-called Yosida formula shows explicitly that the susceptibility
decreases as the number of singlet pairs increases.

\begin{figure}[t]
 \includegraphics[scale=1]{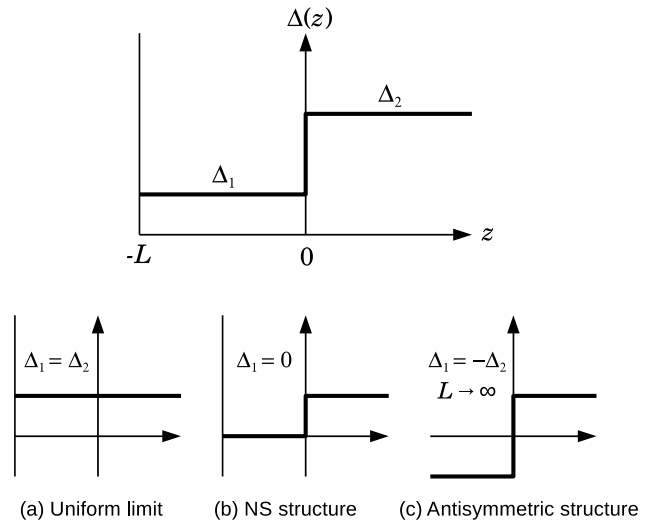} \caption{\label{fig1} Model of
 nonuniform system. Upper panel: a spin-singlet $s$-wave pairing state
 occupying the semi-infinite space $-L < z$ with a specular surface at $z =
 -L$. Lower panel: three particular cases.}
\end{figure}

This paper addresses  how the odd-frequency singlet pairing induced in
nonuniform systems contributes to the superfluid density and the spin
susceptibility. To do that, we consider the following model system that allows
systematic analytical calculation of the physical quantities of interest
here. A singlet $s$-wave pairing state occupies the semi-infinite space $-L
< z$ with a specular surface at $z=-L$ (Fig.\ \ref{fig1}) and is
characterized by the nonuniform gap function
\begin{align}
 \Delta(z) =
  \begin{cases}
   \Delta_1 & (-L < z < 0), \\
   \Delta_2 & (0 < z), \\
  \end{cases}
\end{align}
with $\Delta_1$ and $\Delta_2$ being real constants. The system is assumed
to be clean (impurity free) because the odd-frequency singlet pairs have
odd-parity orbital symmetry and are consequently fragile against impurity
scattering. The gap $\Delta_1$ is treated as a parameter taking values from
$-\Delta_2$ to $\Delta_2$. The case of $\Delta_1 = \Delta_2$ [Fig.\
\ref{fig1}(a)] corresponds to a semi-infinite $s$-wave superconductor with a
uniform gap.  The $s$-wave state is, as is well known, not affected by
surface scattering, so  odd-frequency pairing does not occur in this
case. When $\Delta_1=0$ [Fig.\ \ref{fig1}(b)], the system is analogous to a
normal metal--superconductor (NS) proximity structure with a transparent
interface. It is known that odd-frequency pairing is induced in the N layer
owing to parity mixing at the interface of the NS
structure.\cite{TanakaPRB2007} When the sign of $\Delta_1$ is opposite to
that of $\Delta_2$,  the so-called midgap Andreev bound states
appear around $z=0$.\cite{OhashiJPSJ1996} As was shown in
Ref.~\onlinecite{HigashitaniPRB2012}, the odd-frequency pair amplitude has
a midgap-state pole and there is a close relationship between the midgap
(zero-energy) density of states and the odd-frequency pair amplitude (see
also the Appendix). In the particular case of $\Delta_1 = -\Delta_2$ and $L \to
\infty$ [Fig.\ \ref{fig1}(c)], the pair amplitude at $z=0$ is dominated by
the odd-frequency pairs (see Sec.\ \ref{sec:PA}).

Using the quasiclassical theory of superconductivity,\cite{Serene-Rainer} we can
analyze the pair amplitude, the superfluid density, and the spin
susceptibility in the region $z < 0$ of the above model system. It is shown
that the induced odd-frequency singlet pairing yields a negative pair
density, as in the case of the odd-frequency triplet $s$-wave pairing.  To
investigate the odd-frequency pairing effect on the spin susceptibility, we
generalize the Yosida formula \eqref{Yosida-formula} to the nonuniform
singlet state. The resulting formula describes how the spin susceptibility
is related to the even- and odd-frequency pair amplitudes. It is found from
the generalized Yosida formula that the odd-frequency singlet pairs increase
the spin susceptibility owing to the negative pair density.

 Section\ \ref{sec:QCT}  outlines the framework of the quasiclassical
theory. In Sec.\ \ref{sec:PA}, the quasiclassical theory is applied to the
nonuniform system in Fig.\ \ref{fig1} and  explicit expressions for the
even- and odd-frequency pair amplitudes in $z < 0$ are derived.  The odd-frequency pairing effect on the superfluid density is discussed in Sec.\
\ref{sec:SD}. The Meissner effect in NS proximity
structures is also discussed in this section, with a focus on why the Meissner current is not induced in the
proximity region of a clean N layer with infinitely large layer
width.\cite{HigashitaniJPSJ1995} Finally, the spin susceptibility is
analyzed in Sec.\ \ref{sec:SS}.

\section{Quasiclassical theory}
\label{sec:QCT}

The quasiclassical theory is formulated in terms of a 4 $\times$ 4 matrix
Green's function $\hat{g}(\hat{p},\epsilon,\bm{r})$ in the Nambu space,
where $\hat{p}$ is the unit vector specifying the direction of the Fermi
momentum $\bm{p}_F=\hbar \bm{k}_F$ (and where a spherical Fermi surface is assumed
below), $\epsilon$ is a complex energy variable, and $\bm{r}$ is the spatial
coordinate. The quasiclassical Green's function $\hat{g}$ obeys the
Eilenberger equation
\begin{align}
 i\hbar v_F\hat{p}\cdot\nabla\hat{g}
 + [\hat{\epsilon}(\hat{p},\epsilon,\bm{r}),\ \hat{g}] = 0
 \label{Eilen-eq}
\end{align}
with the normalization condition $\hat{g}^2 = -1$.  In Eq.\
\eqref{Eilen-eq}, $v_F$ is the Fermi velocity and $\hat{\epsilon}$ is an
energy matrix of the form
\begin{align}
 \hat{\epsilon}(\hat{p},\epsilon,\bm{r})
 = \epsilon\hat{\rho}_3  - \hat{v}(\hat{p},\bm{r}) + \hat{\Delta}(\hat{p},\bm{r}),
\end{align}
where $\hat{\rho}_3$ is the third Pauli matrix in particle-hole space,
$\hat{v}$ is a perturbation including Fermi liquid corrections, and
$\hat{\Delta}$ is a mean field (gap function) resulting from Cooper pairing.  In
singlet pairing states, $\hat{\Delta}$ is expressed as
\begin{align}
 \hat{\Delta}(\hat{p},\bm{r}) =
 \begin{bmatrix}
  0 & \Delta(\hat{p},\bm{r})i\sigma_2 \\
  \Delta(\hat{p},\bm{r})^* i\sigma_2 & 0\\
 \end{bmatrix}.
\end{align}
In the present work, the Fermi liquid corrections in $\hat{v}$ are neglected
for simplicity. We can then determine the superfluid density and the spin
susceptibility by calculating the linear response to the spatially uniform
perturbation
\begin{align}
 \hat{v}(\hat{p}) = v_{\hat{p}}\hat{\rho}_3 - h\sigma_3\hat{1},
\end{align}
where $v_{\hat{p}} = p_F\hat{p}\cdot\bm{v}_s$, with $\bm{v}_s$ the superfluid
velocity, $h = \mu_0H$ is the Zeeman coupling of the spin magnetic moment
$\mu_0$ to the external field $H$, and $\hat{1}$ is the unit matrix in
particle-hole space.

In the absence of the perturbation, the 4 $\times$ 4 energy matrix
$\hat{\epsilon}$ for singlet states has the form
\begin{align}
 \hat{\epsilon}
 =
 \begin{bmatrix}
  \epsilon & 0 & 0 & \Delta \\
  0 & \epsilon & -\Delta & 0 \\
  0 & \Delta^* & -\epsilon & 0 \\
  -\Delta^* & 0 & 0 & -\epsilon
 \end{bmatrix}.
\end{align}
The energy matrix $\hat{\epsilon}$ is separated into two $2 \times 2$
subspaces (outer and inner subspaces). The singlet states can therefore be
described by the $2 \times 2$ matrix Eilenberger equation
\begin{align}
 i\hbar v_F\hat{p}\cdot\nabla\hat{g}_{2 \times 2}
 + [\hat{\epsilon}_{2 \times 2}(\hat{p},\epsilon,\bm{r}),\ \hat{g}_{2 \times 2}] = 0
 \label{Eilenberger-eq-2x2}
\end{align}
with
\begin{align}
 \hat{\epsilon}_{2 \times 2}(\hat{p}, \epsilon, \bm{r}) =
 \begin{bmatrix}
  \epsilon & \Delta(\hat{p},\bm{r}) \\
  -\Delta(\hat{p},\bm{r})^* & -\epsilon \\
 \end{bmatrix}.
\end{align}
The perturbation shifts the energy variable $\epsilon$, and the
quasiclassical Green's functions in the outer and inner subspaces are given
by
\begin{align}
 &\hat{g}_{\rm outer} = \hat{g}_{2 \times 2}
 (\hat{p}, \epsilon - v_{\hat{p}} + h, \bm{r}),\\
 &\hat{g}_{\rm inner} = \hat{\rho}_3 \hat{g}_{2 \times 2}
 (\hat{p}, \epsilon - v_{\hat{p}} - h,\bm{r})\hat{\rho}_3.
\end{align}

The Green's function $\hat{g}_{2 \times 2}$ has the matrix structure
\begin{align}
 \hat{g}_{2 \times 2}(\hat{p},\epsilon,\bm{r}) =
 \begin{bmatrix}
  g(\hat{p},\epsilon,\bm{r}) & f(\hat{p},\epsilon,\bm{r}) \\
  -\widetilde{f}(\hat{p},\epsilon,\bm{r}) & -g(\hat{p},\epsilon,\bm{r})
 \end{bmatrix},
\end{align}
where
\begin{align}
 \widetilde{f}(\hat{p}, \epsilon, \bm{r}) = f(-\hat{p}, -\epsilon^*, \bm{r})^*.
\end{align}
The diagonal and off-diagonal elements have the symmetries
\begin{align}
 &g(\hat{p}, \epsilon, \bm{r}) = g(\hat{p}, \epsilon^*, \bm{r})^*,
 \label{symrel-g}\\
 &f(\hat{p}, \epsilon, \bm{r}) = f(-\hat{p}, -\epsilon, \bm{r}).
 \label{symrel-f}
\end{align}

The function $g$ carries information on quasiparticle excitation.  The
local density of states is calculated from $g$ as
\begin{align}
 \begin{split}
  n(\hat{p}, E, \bm{r})
  &= {\rm Im}[g(\hat{p},E+i0,\bm{r})] \\
  &= \frac{1}{2i}[g(\hat{p},E+i0,\bm{r})-g(\hat{p},E-i0,\bm{r})],
 \end{split}
\end{align}
where $E$ is a real energy variable. By using $g$ in the Matsubara representation
($\epsilon = i\epsilon_n$), the supercurrent $\bm{J}(\bm{r})$ and the spin
magnetization $M(\bm{r})$ are obtained from
\begin{gather}
 \bm{J}(\bm{r}) = -2N(0)v_F\frac{\pi}{\beta}\sum_{\epsilon_n}
 \ave{\hat{p}\,g(\hat{p}, i\epsilon_n - v_{\hat{p}}, \bm{r})}_{\hat{p}},
 \label{J-formula}\\
 \frac{M(\bm{r})}{\chi_0^{}H} = 1 - \frac{1}{h}\frac{\pi}{\beta}\sum_{\epsilon_n}
 \sum_{\sigma = \pm}
 \ave{\frac{\sigma}{2}\,g(\hat{p}, i\epsilon_n + \sigma h, \bm{r})}_{\hat{p}},
 \label{M-formula}
\end{gather}
where $N(0)$ is the density of states per spin at the Fermi level in the
normal state and $\chi_0^{} = 2N(0)\mu_0^2$ is the susceptibility in the
normal state.

The function $f$ corresponds to the singlet pair amplitude defined on the
complex $\epsilon$ plane. Equation \eqref{symrel-f} represents a general
symmetry relation for $f$, showing that an even-parity (odd-parity) singlet
pair amplitude has even-frequency (odd-frequency) symmetry.

A more explicit expression for $\hat{g}_{\rm 2 \times 2}$ can be obtained by
expressing it in the form
\begin{align}
 \hat{g}_{\rm 2 \times 2}
 = \frac{2i}{\braket{\bm{r}}{\bm{r}}}\ket{\bm{r}}\bra{\bm{r}} - i,
 \label{g-braket}
\end{align}
where $\ket{\bm{r}}$ and $\bra{\bm{r}}$ are the column and row vectors
satisfying
\begin{align}
 i\hbar v_F\hat{p}\cdot\nabla \ket{\bm{r}}
 &= -\hat{\epsilon}_{2 \times 2}(\hat{p},\epsilon,\bm{r}) \ket{\bm{r}},
 \label{ket-eq}\\
 i\hbar v_F\hat{p}\cdot\nabla \bra{\bm{r}}
 &= \bra{\bm{r}}\hat{\epsilon}_{2 \times 2}(\hat{p},\epsilon,\bm{r}).
 \label{bra-eq}
\end{align}
Noting that $\braket{\bm{r}}{\bm{r}}$ is independent of $\bm{r}$, we can
easily show that $\hat{g}_{\rm 2 \times 2}$ of Eq.\ \eqref{g-braket}
satisfies the Eilenberger equation \eqref{Eilenberger-eq-2x2} with the
normalization condition $\hat{g}_{\rm 2 \times 2}^2 = -1$. The column and
row vectors can be parameterized as
\begin{align}
 \ket{\bm{r}} = a
 \begin{bmatrix} 1 \\ \tf \end{bmatrix},\quad
 \bra{\bm{r}} = a'
 \begin{bmatrix} 1 & \cf \end{bmatrix}.
 \label{braket-ex}
\end{align}
Substitution of Eq.\ \eqref{braket-ex} into Eq.\ \eqref{g-braket} yields the
following parameterization for $\hat{g}_{2 \times 2}$:
\begin{align}
 \hat{g}_{2 \times 2} + i =
 \frac{2i}{1+\cf\tf}
 \begin{bmatrix} 1 \\ \tf \end{bmatrix}
 \begin{bmatrix} 1 & \cf \end{bmatrix}.
 \label{g2x2-ex}
\end{align}
The functions $\cf$ and $\tf$ satisfy the Riccati-type differential
equations
\begin{align}
 i\hbar v_F\hat{p}\cdot\nabla \cf
 &= -2\epsilon \cf + \Delta(\hat{p},\bm{r}) + \Delta(\hat{p},\bm{r})^* \cf^2,
 \label{cf-eq}\\
 i\hbar v_F\hat{p}\cdot\nabla \tf
 &= 2\epsilon \tf + \Delta(\hat{p},\bm{r})^* + \Delta(\hat{p},\bm{r}) \tf^2.
 \label{tf-eq}
\end{align}
In bulk systems with a constant gap function $\Delta(\hat{p})$, the Riccati
equations have solutions
\begin{align}
 &\cf^{\rm bulk}
 = \frac{\Delta(\hat p)}{\epsilon + i\sqrt{|\Delta(\hat{p})|^2 - \epsilon^2}},
 \label{cf-bulk-ex}\\
 &\tf^{\rm bulk}
 = -\frac{\Delta(\hat p)^*}{\epsilon + i\sqrt{|\Delta(\hat{p})|^2 - \epsilon^2}}.
 \label{tf-bulk-ex}
\end{align}
Substituting Eqs.\ \eqref{cf-bulk-ex} and \eqref{tf-bulk-ex} into Eq.\
\eqref{g2x2-ex}, we obtain the well-known bulk solution of $\hat{g}_{2
\times 2}$, i.e.,
\begin{align}
 \hat{g}_{2 \times 2}^{\rm bulk}(\hat{p},\epsilon)
 = \frac{1}{\sqrt{|\Delta(\hat{p})|^2 - \epsilon^2}}
 \begin{bmatrix}
  \epsilon & \Delta(\hat{p}) \\
  -\Delta(\hat{p})^* & -\epsilon \\
 \end{bmatrix}.
\end{align}
One can show from Eqs.\ \eqref{cf-eq}--\eqref{tf-bulk-ex} that $\tf$ is
related to $\cf$ by the transformation
\begin{align}
 \tf(\hat{p}, \epsilon, \bm{r}) = \cf(-\hat{p}, -\epsilon^*, \bm{r})^*.
\end{align}
Moreover, $\cf$ and $\tf$ are found to have the symmetries
\begin{align}
 &\cf(\hat{p},\epsilon,\bm{r}) = {1}/{\cf(\hat{p},\epsilon^*,\bm{r})^*},
 \label{symrel-cf-RA}\\
 &\tf(\hat{p},\epsilon,\bm{r}) = {1}/{\tf(\hat{p},\epsilon^*,\bm{r})^*}.
 \label{symrel-tf-RA}
\end{align}
Equations \eqref{symrel-cf-RA} and \eqref{symrel-tf-RA} can be used to check
the Green's function symmetries of Eqs.\ \eqref{symrel-g} and
\eqref{symrel-f}.

\section{Pair amplitude}
\label{sec:PA}

We now consider the model system in Fig.\ \ref{fig1}. The function $\cf$ obeys
\begin{align}
 i\hbar v_F \hat{p}_z \partial_z \cf = -2\epsilon \cf + \Delta(z)(1 + \cf^2)
 \label{cf-s-wave-eq}
\end{align}
with the following boundary conditions: (i) $\cf \to \cf^{\rm bulk}$ at $z
\to \infty$, (ii) $\cf$ is continuous at $z=0$, and (iii) $\cf$ satisfies
the specular-surface boundary condition $\cf(\hat{p}_z) = \cf(-\hat{p}_z)$
at $z = -L$.

Since $\Delta(z)$ is a real function, $\cf$ has the symmetry
\begin{align}
 \cf(\hat{p}, \epsilon, z) = -\cf(\hat{p}, -\epsilon^*, z)^*.
\end{align}
The corresponding symmetry for the pair amplitude is
\begin{align}
 f(\hat{p}, \epsilon, z) = f(\hat{p}, -\epsilon^*, z)^*.
 \label{symrel-TRI-f}
\end{align}
This shows that $f$ for $\epsilon=i\epsilon_n$ is a real quantity.

In general, the pair amplitude has even-frequency (EF) and odd-frequency
(OF) components,
\begin{align}
 f(\hat{p}, \epsilon, z)
 = f^{\rm EF}(\hat{p}, \epsilon, z) + f^{\rm OF}(\hat{p}, \epsilon, z).
 \label{f-EF-OF}
\end{align}
Combining the symmetries \eqref{symrel-TRI-f} and \eqref{symrel-f}, we find
that $\widetilde{f}$ can be decomposed as
\begin{align}
 \widetilde{f}(\hat{p}, \epsilon, z)
 = f^{\rm EF}(\hat{p}, \epsilon, z) - f^{\rm OF}(\hat{p}, \epsilon, z).
 \label{tf-EF-OF}
\end{align}

In the model system, we can solve analytically the Riccati equation
\eqref{cf-s-wave-eq}. The general solution can be written in the form
\begin{align}
 &\cf(z < 0) = \frac{\cf_1 + C_1 e^{-s_p\kappa_1 z}}
 {1 + \cf_1 C_1 e^{-s_p\kappa_1 z}},\\
 &\cf(z > 0) = \frac{\cf_2 + C_2 e^{-s_p\kappa_2 z}}
 {1 + \cf_2 C_2 e^{-s_p\kappa_2 z}},
\end{align}
where
\begin{align}
 &\cf_i = \frac{\Delta_i}{\epsilon + i\sqrt{\Delta_{i}^2 - \epsilon^2}}
 \ \ (i=1,2),\\
 &\kappa_i = \frac{2\sqrt{\Delta_{i}^2 - \epsilon^2}}{\hbar v_F|\hat{p}_z|},
 \\
 &s_p = {\rm sgn}(\hat{p}_z),
\end{align}
and $C_i$s are constants to be determined from the boundary conditions.
Imposing the boundary conditions, we obtain
\begin{align}
 \cf(z<0)
 &= \frac{1+s_p}{2}\frac{\cf_1 + C e^{-\kappa_1(z+2L)}}{1 + \cf_1 C e^{-\kappa_1(z+2L)}}
 \notag\\
 &+ \frac{1 - s_p}{2}\frac{\cf_1 + C e^{\kappa_1 z}}{1 + \cf_1 C e^{\kappa_1 z}},
 \label{cf1-ex}\\
 \cf(z>0)
 &= \frac{1+s_p}{2}\frac{\cf_2 - C' e^{-\kappa_2 z}}{1 - \cf_2 C' e^{-\kappa_2 z}}
 \notag\\
 &+ \frac{1 - s_p}{2}\cf_2,
 \label{cf2-ex}
\end{align}
where
\begin{align}
 C = \frac{\cf_2-\cf_1}{1 - \cf_1\cf_2},\quad
 C' = \frac{C(1 - e^{-2\kappa_1L})}{1 - C^2e^{-2\kappa_1L}}.
\end{align}
The factors $(1 + s_p)/2$ and $(1 - s_p)/2$ in Eqs.\ \eqref{cf1-ex} and
\eqref{cf2-ex} select $\cf$ with $\hat{p}_z > 0$ and with $\hat{p}_z < 0$,
respectively.

In what follows, we shall focus on the region $z < 0$. Using Eq.\ \eqref{cf1-ex},
we find that the quasiclassical Green's function $\hat{g}_{2 \times 2}$ in $z<0$ is  given as
\begin{widetext}
\begin{align}
 \hat{g}_{2 \times 2}(z<0) + i
 &= \frac{(1 + s_p)i}{D}
 \biggl(
 \begin{bmatrix} 1 \\ -\cf_1 \end{bmatrix}
 + C e^{\kappa_1z}
 \begin{bmatrix} \cf_1 \\ -1 \end{bmatrix}
 \biggr)
 \biggl(
 \begin{bmatrix} 1 & \cf_1 \end{bmatrix}
 + Ce^{-\kappa_1(z+2L)}\begin{bmatrix} \cf_1 & 1 \end{bmatrix}
 \biggr)
 \notag\\
 &+ \frac{(1 - s_p)i}{D}
 \biggl(
 \begin{bmatrix} 1 \\ -\cf_1 \end{bmatrix}
 + C e^{-\kappa_1(z+2L)}
 \begin{bmatrix} \cf_1 \\ -1 \end{bmatrix}
 \biggr)
 \biggl(
 \begin{bmatrix} 1 & \cf_1 \end{bmatrix}
 + Ce^{\kappa_1z}\begin{bmatrix} \cf_1 & 1 \end{bmatrix}
 \biggr)
 \label{g2x2-1-ex}
\end{align}
with $D = (1 - \cf_1^2)(1 - C^2e^{-2\kappa_1L})$.

Equation \eqref{g2x2-1-ex} depends on $\Delta_2$ via the constant $C$. When
$\Delta_1 = \Delta_2$ (the uniform limit), $C$ vanishes and then Eq.\
\eqref{g2x2-1-ex} is reduced to the bulk solution, as expected from the fact
that the $s$-wave pairing state is not affected by surface
scattering. However, the spatial inhomogeneity arising from  $\Delta_1 \neq
\Delta_2$ makes $C$ finite. For example, in the NS structure, we have $\cf_1
= 0$ for $\epsilon = E+i0$ and then $C = \cf_2$. Note that, in this case,
Eq.\ \eqref{g2x2-1-ex} for $\hat{p}_z>0$ and $L \to \infty$ can be expressed
in the form
\begin{align}
 \hat{g}_{2 \times 2}(z<0) + i
 = 2i\hat{\rho}_3
 \left(
 \begin{bmatrix} 1 \\ 0 \end{bmatrix}
 e^{iqz} + \cf_2
 \begin{bmatrix} 0 \\ 1 \end{bmatrix}
 e^{-iqz}
 \right)
 \begin{bmatrix} 1 & 0 \end{bmatrix}
 e^{-iqz},
 \label{g2x2-NS-ex}
\end{align}
where $q = E/\hbar v_F|\hat{p}_z|$. The two column vectors on the right-hand
side of Eq.\ \eqref{g2x2-NS-ex} represent the Andreev scattering process in
N of the NS structure. This shows that $\cf_2$ for real energies gives the
Andreev reflection amplitude.

The upper-right matrix element of Eq.\ \eqref{g2x2-1-ex} gives the pair
amplitude $f$ in $z<0$. In the expression for $f$, the terms $\propto s_p$
are odd-parity pair amplitudes and therefore have OF symmetry. We can check
the frequency symmetry using the relation $\cf_i(-\epsilon) =
-\cf_i^{-1}(\epsilon)$ ($i=1,2$). We thus find that in the region $z<0$
there coexist  EF and OF pairs with  amplitudes
\begin{align}
 &f^{\rm EF}(\hat{p},\epsilon,z<0) = i\,
 \frac{2\cf_1\left(1 + C^2e^{-2\kappa_1L}\right)
 + (1+\cf_1^2)C\left(e^{\kappa_1z}+e^{-\kappa_1(z+2L)}\right)}
 {(1-\cf_1^2)\left(1-C^2e^{-2\kappa_1L}\right)},
 \label{f-EF-ex}\\
 &f^{\rm OF}(\hat{p},\epsilon,z<0) = -s_pi\,
 \frac{C \left(e^{\kappa_1z} - e^{-\kappa_1(z+2L)}\right)}{1 - C^2e^{-2\kappa_1L}},
 \label{f-OF-ex}
\end{align}
\end{widetext}
respectively. The OF pair amplitude is proportional to $C$. This means that it vanishes in
the uniform limit and then the EF pair amplitude takes the bulk form $f^{\rm
bulk} = \Delta_1/\sqrt{\Delta_1^2 - \epsilon^2}$.

When $\Delta_1 = 0$ (NS structure), the EF and OF pairs for $\epsilon =
E+i0$ have  amplitudes
\begin{align}
 &f^{\rm EF}(\hat{p},\epsilon,z<0)
 = i\,\frac{\cf_2\left(e^{\kappa_1z} + e^{-\kappa_1(z+2L)}\right)}
 {1 - \cf_2^2e^{-2\kappa_1L}},
 \label{f-EF-N}\\
 &f^{\rm OF}(\hat{p},\epsilon,z<0)
 = -s_pi\,\frac{\cf_2\left(e^{\kappa_1z} - e^{-\kappa_1(z+2L)}\right)}
 {1 - \cf_2^2e^{-2\kappa_1L}},
 \label{f-OF-N}
\end{align}
respectively. The pair amplitudes are proportional to the Andreev reflection amplitude
$\cf_2$. The denominator with $\cf_2$ describes the multiple Andreev
scattering effect in an N layer of finite width $L$.  Equations
\eqref{f-EF-N} and \eqref{f-OF-N} can also be applied to the case of
$\epsilon = i\epsilon_n$ with $\epsilon_n>0$. Then, the spatial dependence
of the pair amplitudes is characterized by
\begin{align}
 \kappa_1
 = \frac{2|\epsilon_n|}{\hbar v_F |\hat{p}_z|}
 = \frac{|2n+1|}{\xi_N(T)|\hat{p}_z|}
\end{align}
with $\xi_N(T) = \hbar v_F / 2\pi k_BT$ being the coherence length in the N
layer. The Matsubara pair amplitudes in the N layer decay exponentially from
$z=0$ and penetrate to a distance $\sim \xi_N(T)$. The EF and OF pair
amplitudes have the same magnitude in the limit $L/\xi_N(T) \gg 1$. This is
because in that limit the total propagator $f$ with $\hat{p}_z > 0$ does not
carry  information on the proximity effect, i.e., $f(\hat{p}, \epsilon,
z<0) = 0$ for $\hat{p}_z > 0$.

Let us consider infinite systems with $\Delta_1 \neq 0$. Taking the limit $L
\to \infty$ in Eqs.\ \eqref{f-EF-ex} and \eqref{f-OF-ex}, we obtain
\begin{align}
 &f^{\rm EF}(\hat{p},\epsilon,z<0) = i\,
 \frac{2\cf_1 + (1+\cf_1^2)Ce^{\kappa_1z}}{1-\cf_1^2},
 \label{f-EF-Linf}
 \\
 &f^{\rm OF}(\hat{p},\epsilon,z<0) = -s_pi\,Ce^{\kappa_1z}.
 \label{f-OF-Linf}
\end{align}
It should be noted here that $C$ diverges at $\epsilon=0$ when ${\rm
sgn}(\Delta_1\Delta_2) < 0$. This corresponds to the pole of the midgap
Andreev bound states localized around $z=0$. The OF pair amplitude has the
midgap-state pole, whereas the EF pair amplitude does not, because $1+\cf_1^2
\propto \epsilon$ in the low-energy limit. As shall be shown in the Appendix,
the midgap (zero-energy) density of states can be written in terms of the OF
pair amplitude.

In the particular case of $L \to \infty$ and $\Delta_1 = -\Delta_2$
(antisymmetric structure), we get from Eqs.\ \eqref{f-EF-Linf} and
\eqref{f-OF-Linf} the following explicit expressions for the EF and OF pair
amplitudes:
\begin{align}
 f^{\rm EF}(\hat{p},\epsilon,z<0)
 &= \frac{\Delta_1}{\sqrt{\Delta_1^2-\epsilon^2}}(1-e^{\kappa_1z}),
 \label{f-EF-antisymmetric}\\
 f^{\rm OF}(\hat{p},\epsilon,z<0)
 &= s_p\frac{i\Delta_1}{\epsilon}e^{\kappa_1z}.
 \label{f-OF-antisymmetric}
\end{align}
In this case, the total pair amplitude at $z = 0$ is dominated by the OF
pairs.

\section{Superfluid density}
\label{sec:SD}

In the system considered above, supercurrent can flow along the surface
(perpendicular to the $z$ axis). The corresponding superfluid density can be
obtained by calculating the linear response of $g$ to $v_{\hat{p}} =
p_F\hat{p}_xv_s$.  The linear deviation $\delta g$ of the Matsubara $g$
function is given as
\begin{align}
 \delta g(\hat{p},i\epsilon_n-v_{\hat{p}},z) = -v_{\hat{p}}\,g'(\hat{p},i\epsilon_n,z)
 \label{dg-vp-rel}
\end{align}
with
\begin{align}
 g'(\hat{p}, i\epsilon_n, z) = (-i)
 \frac{\partial}{\partial \epsilon_n}g(\hat{p},i\epsilon_n,z).
 \label{gp-g-rel}
\end{align}

Equation \eqref{gp-g-rel} relates explicitly the response function $g'$ to
the unperturbed Green's function $g$. Such a definition of $g'$ is, however,
not so convenient for the analysis of the Cooper pairing effect on the
superfluid density. A more useful formula can be obtained by starting with
Eq.\ \eqref{g2x2-ex}, giving the expression
\begin{align}
 g + i = \frac{2i}{1 + \cf\tf}.
 \label{g-cf-ex}
\end{align}
Let $\delta\cf$ be the linear deviation of $\cf$. Replacing $\cf$ in Eq.\
\eqref{g-cf-ex} by $\cf + \delta\cf$, we obtain
\begin{align}
 \delta g = -v_{\hat{p}}\,g'
 = -\frac{2i}{(1+\cf\tf)^2}(\delta\cf\tf + \cf\delta\tf).
\end{align}
Moreover, using the expression
\begin{align}
 f = \frac{2i\cf}{1+\cf\tf}
\end{align}
for the pair amplitude, we get the following formula for the response
function:
\begin{align}
 g'(\hat{p},i\epsilon_n,z) = \Lambda(\hat{p},i\epsilon_n,z)
 f(\hat{p},i\epsilon_n,z)\widetilde{f}(\hat{p},i\epsilon_n,z)
 \label{gp-ff-rel}
\end{align}
with
\begin{align}
 \Lambda(\hat{p},i\epsilon_n,z) = -\frac{1}{2iv_{\hat{p}}}
 \left\{\frac{\delta\cf}{\cf} +
 \frac{\delta\tf}{\tf}\right\}{(\hat{p},i\epsilon_n,z)}.
 \label{Lambda-def}
\end{align}
In Eq.\ \eqref{Lambda-def}, the notation
$\{\cdots\}{(\hat{p},i\epsilon_n,z)}$ denotes that all the functions in the
curly braces have the same argument $(\hat{p},i\epsilon_n,z)$.

From Eqs.\ \eqref{gp-ff-rel}, \eqref{dg-vp-rel}, and \eqref{J-formula}, we
find that the superfluid fraction is given by
\begin{align}
 \frac{n_s(z)}{n} = \frac{\pi}{\beta}\sum_{\epsilon_n}\ave{3\hat{p}_x^2
 \left\{\Lambda P\right\}{(\hat{p},i\epsilon_n,z)}}_{\hat{p}}
 \label{ns-formula}
\end{align}
with
\begin{align}
 P = f\widetilde{f} = (f^{\rm EF})^2 - (f^{\rm OF})^2.
 \label{P-def}
\end{align}
Since $f$ with $\epsilon=i\epsilon_n$ is a real function,
$P(\hat{p},i\epsilon_n,z)$ is a real quantity. The function
$\Lambda(\hat{p},i\epsilon_n,z)$ is also a real quantity because $\cf$ and
$\delta\cf$ in the Matsubara representation are purely imaginary and real,
respectively.  Moreover, one can show that $\Lambda$ has the symmetry
\begin{align}
 \Lambda(\hat{p},i\epsilon_n,z)
 = \Lambda(-\hat{p},i\epsilon_n,z)
 = \Lambda(\hat{p},-i\epsilon_n,z).
\end{align}
Namely, $\Lambda$ is even in $\hat{p}$ and in $\epsilon_n$.

It is instructive to compare Eq.\ \eqref{ns-formula} with the corresponding
formula for a dirty singlet superconductor, i.e., Eq.\
\eqref{Usadel-J-formula}. The superfluid fraction in the dirty system is
obtained by the replacement
\begin{align}
 \Lambda \to 2\tau_{\rm tr}/\hbar,\quad P \to (f^{\rm EF}_{\rm SW})^2,
\end{align}
where $f^{\rm EF}_{\rm SW}$ denotes the even-frequency $s$-wave pair
amplitude. Note that $\hbar v_F\Lambda/2$ coincides with the mean free path
$v_F\tau_{\rm tr}$.  This implies that the quantity $\hbar v_F\Lambda/2$
corresponds to the range of the linear response kernel; in other words,
$n_s(z)$ is determined depending only on $\bm{v}_s$ in the region of width
$\sim \hbar v_F\Lambda/2$ around position $z$. The pair density $P$ in the
dirty singlet superconductor does not contain the OF pair amplitude. This is
because impurity scattering destroys non-$s$-wave pairs and singlet $s$-wave
pairing has even-frequency symmetry.

However, in the clean systems under consideration, the OF pairs
exist except at the uniform limit. Equation \eqref{P-def} shows that the OF
pairing yields a {\em negative} pair density.

The rest of this section is devoted to a discussion of the superfluid density in the three
particular clean system cases: the uniform limit [Fig.\
\ref{fig1}(a)], the NS structure [Fig.\ \ref{fig1}(b)], and the
antisymmetric structure [Fig.\ \ref{fig1}(c)].

\subsection{Uniform limit}

In the case of $\Delta_1 = \Delta_2 \equiv \Delta$, we have
\begin{align}
 \cf = \cf^{\rm bulk},\quad f^{\rm EF} = f^{\rm bulk},\quad f^{\rm OF} = 0.
\end{align}
Using
\begin{align}
 \frac{\partial\cf^{\rm bulk}}{\partial\epsilon_n}
 = \frac{\partial}{\partial\epsilon_n}
 \frac{-i\Delta}{\epsilon_n + \sqrt{\Delta^2 + \epsilon_n^2}}
 = -\frac{\cf^{\rm bulk}}{\sqrt{\Delta^2+\epsilon_n^2}},
\end{align}
we can obtain
\begin{align}
 \Lambda = \frac{1}{\sqrt{\Delta^2 + \epsilon_n^2}}.
 \label{Lambda-ex-HL}
\end{align}
Note that $\hbar v_F \Lambda/2$ coincides with the $\epsilon_n$-dependent
coherence length $\xi(\epsilon_n, \Delta) = \hbar v_F /
2\sqrt{\Delta^2+\epsilon_n^2}$\,, which determines the range of the linear
response kernel in the clean superconductor with gap $\Delta$. The pair
density in the uniform superconductor is
\begin{align}
 P = (f^{\rm bulk})^2 = \frac{\Delta^2}{\Delta^2 + \epsilon_n^2}.
 \label{P-ex-HL}
\end{align}
Substitution of Eqs.\ \eqref{Lambda-ex-HL} and \eqref{P-ex-HL} into Eq.\
\eqref{ns-formula} leads to
\begin{align}
 \frac{n_s}{n} = \frac{\pi}{\beta}\sum_{\epsilon_n}
 \frac{\Delta^2}{(\Delta^2+\epsilon_n^2)^{3/2}},
 \label{ns-bulk}
\end{align}
which is the well-known result for the superfluid fraction in clean bulk
$s$-wave superconductors.

\subsection{NS structure}

This subsection focuses on the N layer of the NS structure.

In the clean N layer, the superfluid density is known to take a spatially
constant value despite the existence of the spatially varying pair
amplitude. This property can be readily shown from the Eilenberger equation
in the normal state,
\begin{align}
 i\hbar v_F \hat{p}_z \partial_z \hat{g}_{2 \times 2}^N
 + [(i\epsilon_n - v_{\hat{p}})\hat{\rho}_3,\ \hat{g}_{2 \times 2}^N] = 0.
\end{align}
The spatial dependence of $\hat{g}_{2 \times 2}^N(z)$ is described by
\begin{align}
 \hat{g}_{2 \times 2}^N(z)
 = e^{-\kappa_N(z-z')\hat{\rho}_3} \hat{g}_{2 \times 2}^N(z') e^{\kappa_N(z-z')\hat{\rho}_3}
\end{align}
with $\kappa_N = (\epsilon_n + iv_{\hat{p}})/\hbar v_F \hat{p}_z$. It
follows that the diagonal element $g^N(z)$ of $\hat{g}_{2 \times 2}^N(z)$ is
independent of $z$. This also means that the Meissner response of the clean
N layer is completely nonlocal.\cite{HigashitaniJPSJ1995}

From the $n_s$ formula \eqref{ns-formula}, the superfluid density in the N
layer is obtained as follows. The function $\Lambda$ is determined from Eq.\
\eqref{cf1-ex} with $\cf_1 = 0$ and $\kappa_1 = 2|\epsilon_n|/\hbar v_F
|\hat{p}_z|$. The result is
\begin{align}
 \Lambda = \frac{2L}{\hbar v_F |\hat{p}_z|}
 + \frac{1}{\sqrt{\Delta_2^2 + \epsilon_n^2}}.
\end{align}
The first term is proportional to $L$ because of the nonlocal response of
the clean N layer. The second term implies that the N-side superfluid
density includes  information on $\bm{v}_s$ in the region of $0 < z \alt
\xi(\epsilon_n, \Delta_2)$ on the S side. The pair density $P$ in the N
layer is obtained from Eqs.\ \eqref{f-EF-N} and \eqref{f-OF-N} as
\begin{align}
 P = (f^{\rm EF})^2 - (f^{\rm OF})^2 = \left|\frac{2\cf_2}
 {1 - \cf_2^2e^{-2\kappa_1L}}\right|^2e^{-2\kappa_1L}.
\end{align}
The spatially dependent terms in $(f^{\rm EF})^2$ and $(f^{\rm OF})^2$
cancel out in $P$.

\begin{figure}[t]
 \includegraphics[width=6cm]{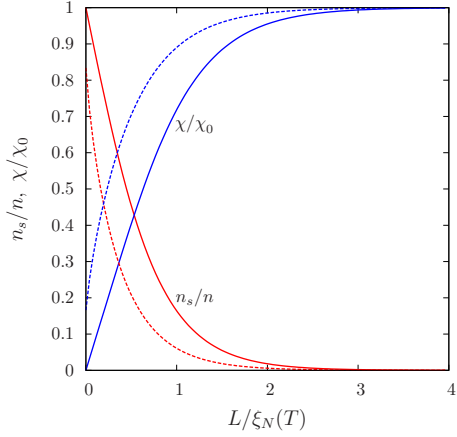}
 \caption{\label{fig2} (Color online) $n_s/n$ and $\chi/\chi_0^{}$ in the
 clean N layer of the NS structure as a function of $L/\xi_N(T)$. The solid
 lines are the results for $T/T_c = 0.1$ and the dashed lines are for $T/T_c
 = 0.5$ (where $T_c$ is the transition temperature of the superconductor).}
\end{figure}

In Fig.\ \ref{fig2}, the superfluid fraction $n_s/n$ in the N layer is
plotted as a function of $L$ scaled by $\xi_N(T)$ at given reduced
temperatures $T/T_c = 0.1$ and 0.5. In the limit $L \to 0$, the superfluid
fraction coincides with that in the uniform state with gap $\Delta_2$. As
$L$ increases, $n_s/n$ decreases and becomes exponentially small for $L \gg
\xi_N(T)$.  In the limit $L/\xi_N(T) \to \infty$, the superfluid density
vanishes because $|f^{\rm OF}|$ is then equal to $|f^{\rm EF}|$ and
consequently $P=0$.

The vanishing superfluid density in the clean N layer for $L/\xi_N(T) \gg 1$
has been noted in a study of the Meissner effect in NS
structures.\cite{HigashitaniJPSJ1995} Mathematically, this is a consequence
of $g^N(z)$ being constant and  therefore being equal to the normal-state
value everywhere in an N layer of infinitely large layer width. The
question, however, remains as to why  supercurrent does not flow even in the
proximity region with a finite pair amplitude.  The present theory provides the
following answer: because the pair density associated with OF pairing is
negative, the supercurrent carried by OF pairs flows in the opposite direction
to $\bm{v}_s$ and compensates for the conventional supercurrent carried by EF
pairs.

In dirty systems, in contrast, the OF singlet pairs are destroyed by
impurity scattering. As a result, the dirty N layer exhibits a
(diamagnetic) Meissner effect similar to that in conventional
superconductors. The diamagnetic Meissner current also flows in the clean N
layer when $L/\xi_N(T) \alt 1$.  In this case, imbalance between $|f^{\rm
EF}|$ and $|f^{\rm OF}|$ ($|f^{\rm EF}| > |f^{\rm OF}|$) is caused by
surface scattering.

\subsection{Antisymmetric  structure}
\label{sec:SD:ASS}

\begin{figure}[t]
 \includegraphics[width=8cm]{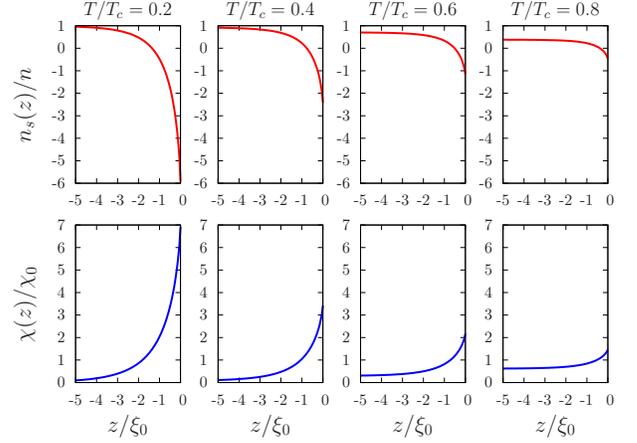}
 \caption{\label{fig3} (Color online) Spatial dependence of $n_s(z<0)/n$ and
 $\chi(z<0)/\chi_0^{}$ in the antisymmetric structure at $T/T_c = 0.2$, 0.4,
 0.6, and 0.8.}
\end{figure}

We now turn to the antisymmetric structure. Plotted in the upper part of Fig.\
\ref{fig3} is  $n_s(z<0)/n$ in the antisymmetric structure as a function
of $z/\xi_0$, where $\xi_0 = \hbar v_F / 2\pi k_B T_c$.  The superfluid
density $n_s(z)$ depends strongly on position $z$, unlike that in the N
layer of the NS system. In the antisymmetric structure, $n_s(z)$ has the
bulk value in the region $z \ll -\xi_0$ but takes a negative value around $z
= 0$. The magnitude of the negative superfluid density at $z=0$ increases
with decreasing temperature $T$.

Let us discuss $n_s(z=0)$. The function $\Lambda$ at $z=0$ has the form
\begin{align}
 \Lambda(z=0)
 = \frac{1}{2}\sum_{i=1,2}\frac{1}{\sqrt{\Delta_i^2+\epsilon_n^2}}
 = \frac{1}{\sqrt{\Delta_1^2+\epsilon_n^2}}.
\end{align}
The pair density at $z=0$ is dominated by the OF pairs and takes the
negative value
\begin{align}
 P(z=0) = -\frac{\Delta_1^2}{\epsilon_n^2}.
\end{align}
The resulting superfluid density at $z=0$ is
\begin{align}
 \frac{n_s(0)}{n} =
 -\frac{\pi}{\beta}\sum_{\epsilon_n}\frac{\Delta_1^2}{\epsilon_n^2
 \sqrt{\Delta_1^2 + \epsilon_n^2}} < 0.
 \label{negative-ns-antisymmetric}
\end{align}

\begin{figure}[t]
 \includegraphics[width=8cm]{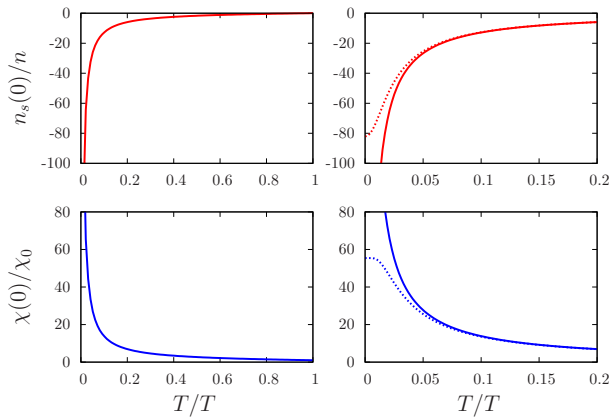}
 \caption{\label{fig4} (Color online) Temperature dependence of $n_s(z)/n$
 and $\chi(z)/\chi_0^{}$ at $z=0$ in the antisymmetric structure.  In the
 calculations, $\Delta_1$ is assumed to have the same temperature dependence
 as that of the bulk $s$-wave gap. The solid lines are the results from
 linear response theory. The dotted lines are the full numerical results
 obtained from the current formula \eqref{J-formula} with $p_Fv_s/k_BT_c =
 0.05$ and the magnetization formula \eqref{M-formula} with $\mu_0H/k_BT_c =
 0.05$. The right panels demonstrate the breakdown of  linear response
 theory at low temperatures below $\sim \negthickspace p_Fv_s/k_B$ or $\sim \negthickspace\mu_0H/k_B$.}
\end{figure}

Equation \eqref{negative-ns-antisymmetric} predicts the temperature
dependence of $n_s(0)/n$ as shown in the upper-left panel of Fig.\
\ref{fig4}. The superfluid fraction has a large negative value at low
temperatures and diverges in the $T \to 0$ limit. It is obvious that the
low-temperature divergence is due to the midgap-state pole of the OF pair
amplitude $f^{\rm OF}$. Strictly, however, $n_s(0)/n$ does not
diverge. The divergence is due to the breakdown of  linear response
theory at low temperatures. To demonstrate this, the low-temperature
behavior of $J(z=0)/nv_s$ calculated from the general current formula
\eqref{J-formula} with $p_Fv_s/k_BT_c = 0.05$ is plotted with a dotted line
in the upper-right panel of Fig.\ \ref{fig4}. The dotted line deviates from
the linear response result (solid line) below $T/T_c \sim p_Fv_s/k_BT_c$.

The origin of the deviation can be understood by expressing Eq.\
\eqref{J-formula} in terms of the local density of states, $n(\hat{p}, E,
z)$,
\begin{align}
 J(z) = 2N(0)v_F\int_{-\infty}^{\infty}dE\,n_F(E)\ave{\hat{p}_x\,
 n(\hat{p},E_{\hat{p}},z)}_{\hat{p}}.
 \label{exact-J}
\end{align}
Here, $E_{\hat{p}} = E - v_{\hat{p}}$ and $n_F(E) = 1/(e^{\beta E}+1)$ is
the Fermi function. The midgap-state pole of $f^{\rm OF}$ yields the
zero-energy peak in the local density of states at $z=0$ (see the Appendix),
\begin{align}
 n_{\rm midgap}(\hat{p}, E, 0) = \pi|\Delta_1|\delta(E).
\end{align}
The contribution from the midgap states to $J(0)$ is evaluated to be
\begin{align}
 J_{\rm midgap}(0) = 2N(0)v_F\pi|\Delta_1|\ave{\hat{p}_xn_F(v_{\hat{p}})}_{\hat{p}}.
 \label{Jmidgap-nF-ex}
\end{align}
Taking the $T \to 0$ limit, we obtain
\begin{align}
 J_{\rm midgap}(0) \xrightarrow{T \to 0} -\frac{1}{2}N(0)v_F\pi|\Delta_1|
 = -n\frac{3\pi|\Delta_1|}{4p_F}.
 \label{Jmidgap-ex}
\end{align}
This result is independent of $v_s$ and suggests the breakdown of  linear
response theory. Since $n_F(v_{\hat{p}})$ in Eq.\ \eqref{Jmidgap-nF-ex} cannot
be expanded in powers of $v_{\hat{p}}$ at low temperatures $k_BT \alt
p_Fv_s$,  linear response theory does not give the correct value of the
midgap-state current at $T=0$.  Linear response theory can still be used
to evaluate the contribution from continuum states. At $T=0$, the continuum
states carry the supercurrent $nv_s$. Adding the two contributions, we find
that
\begin{align}
 J(0) \xrightarrow{T \to 0} n\left(v_s - \frac{3\pi|\Delta_1|}{4p_F}\right).
\end{align}
We see that the superfluid fraction defined as $J(0)/nv_s$ takes the
zero-temperature value
\[
1 - \frac{3\pi|\Delta_1|}{4p_Fv_s} \simeq - \frac{3\pi|\Delta_1|}{4p_Fv_s},
\]
which is finite, though it has a large negative value for small $v_s$, as
suggested by  linear response theory.

Equation \eqref{Jmidgap-ex} shows that the magnitude of $J_{\rm midgap}$ is
as large as the critical current density $\sim n|\Delta_1|/p_F$, as was
noted in Ref.\ \onlinecite{HigashitaniJPSJ1997}. The fact that the
midgap states carry such a large current can be understood as follows. Since
the antisymmetric structure has one midgap state for each parallel momentum
$\bm{p}_\| = (p_x, p_y)$, the magnitude of the total midgap-state current
$\int dz\,J_{\rm midgap}(z)$ is of the order of $k_F^2v_F$. The midgap
states are localized in the region $|z| \sim \hbar v_F / |\Delta_1|$. Hence,
$J_{\rm midgap} \sim k_F^2v_F/(\hbar v_F/|\Delta_1|) \sim n |\Delta_1|/p_F$.

\section{Spin susceptibility}
\label{sec:SS}

The spin susceptibility can be calculated from
\begin{align}
 \delta g(\hat{p}, i\epsilon_n + \sigma h, z)
 = \sigma h\, g'(\hat{p},i\epsilon_n,z).
 \label{dg-h-ex}
\end{align}
From Eqs.\ \eqref{dg-h-ex}, \eqref{gp-ff-rel}, and \eqref{M-formula}, we
obtain the following formula for the local susceptibility $\chi(z) =
M(z)/H$:
\begin{align}
 \frac{\chi(z)}{\chi_0^{}} = 1 - \frac{\pi}{\beta}\sum_{\epsilon_n}
 \ave{\{\Lambda P\}(\hat{p},i\epsilon_n,z)}_{\hat{p}}.
 \label{generalized-Yosida-formula}
\end{align}
Equation \eqref{generalized-Yosida-formula} provides a natural
generalization of the Yosida formula \eqref{Yosida-formula} to the
nonuniform $s$-wave state. It is found from the generalized Yosida formula
that the OF pairing gives an anomalous contribution to the susceptibility:
since OF pairing yields a negative pair density, it enhances the
susceptibility even though its spin symmetry is singlet.

As with the superfluid density, the susceptibility in the N layer of the NS
structure is independent of $z$, and its value strongly depends on the layer
width $L$ (Fig.\ \ref{fig2}). With increasing $L$ from $0$ to $\infty$, the
susceptibility in the N layer changes from the bulk $s$-wave-state value
$\chi^{\rm bulk}$ to the normal-state value $\chi_0^{}$. The saturation to
$\chi_0^{}$ in the $L \to \infty$ limit reflects the fact that $|f^{\rm
EF}|$ and $|f^{\rm OF}|$ become equal to each other in that limit.

In the antisymmetric structure, the OF pairing causes substantial
susceptibility enhancement at $z=0$ (Fig.\ \ref{fig3}). Linear response
theory in this case gives
\begin{align}
 \frac{\chi(0)}{\chi_0^{}} = 1 + \frac{\pi}{\beta}\sum_{\epsilon_n}
 \frac{\Delta_1^2}{\epsilon_n^2\sqrt{\Delta_1^2+\epsilon_n^2}} > 1.
\end{align}
With decreasing $T$ from $T_c$, the normalized susceptibility
$\chi(0)/\chi_0$ {\em increases} from unity and diverges in the
zero-temperature limit (Fig.\ \ref{fig4}).

As in the case of the superfluid density, the
susceptibility divergence results from the failure of  linear response theory to evaluate
correctly the contribution from the midgap states at low temperatures. In
the temperature dependence of the susceptibility, the deviation from the
full theory occurs below $T \alt \mu_0H/k_B$, as demonstrated for
$\mu_0H/k_BT_c = 0.05$ in the lower-right panel of Fig.\ \ref{fig4}. The
correct low-temperature behavior can be obtained from Eq.\ \eqref{M-formula}
or, equivalently, from
\begin{align}
 \!\!\frac{M(z)}{\chi_0^{}H} = 1 + \frac{1}{h}\int_{-\infty}^{\infty} dE\,n_F(E)
 \sum_{\sigma=\pm}\frac{\sigma}{2}\ave{n(\hat{p}, E_\sigma, z)}_{\hat{p}}
 \label{M0-dos-rel}
\end{align}
with $E_\sigma = E + \sigma h$. The contribution from the midgap states to
$M(0)/\chi_0^{}H$ is
\begin{align}
 \frac{M_{\rm midgap}(0)}{\chi_0^{}H}
 = \frac{\pi|\Delta_1|}{2h}\tanh\frac{\beta h}{2}.
 \label{Mmidgap-ex}
\end{align}
This result implies the breakdown of  linear response theory at low
temperatures with $\beta h = \mu_0H/k_BT\agt 1$.  Equation
\eqref{Mmidgap-ex} also implies that the magnitude of the total midgap-state
magnetization $\sim M_{\rm midgap}(0) \times \hbar v_F/|\Delta_1|$ at
low temperature is of the order of $k_F^2 \mu_0$, in which the factor
$k_F^2$ originates from the number of  midgap states.

At $T=0$, the continuum states give the contribution $-1$, which cancels out
the first term in Eq.\ \eqref{M0-dos-rel}. It follows that
\begin{align}
 \frac{M(0)}{\chi_0^{}H} = \frac{\chi(0)}{\chi_0^{}}
 \xrightarrow{T \to 0} \frac{\pi|\Delta_1|}{2h}.
 \label{M-T0-ex}
\end{align}
The zero-temperature susceptibility is inversely proportional to $H$ and
takes a large positive value for small $H$.

\begin{acknowledgments}
 I  thank Y.~Asano for useful discussions.  This work was
 supported in part by the ``Topological Quantum Phenomena'' (Grant No.\ 22103003)
 KAKENHI on Innovative Areas from MEXT of Japan.
\end{acknowledgments}

\appendix*

\section{Odd-frequency pairing and the zero-energy density of states}

The purpose of this appendix is to show that the zero-energy density of
states can be obtained from the odd-frequency pair amplitude. The system
considered here is similar to that in the upper panel in Fig.\ \ref{fig1},
but here we do not assume a specific profile of $\Delta(z)$ except that
$\Delta(z)$ takes an asymptotic constant value $\Delta_2$ at $z \to \infty$.

We start with the following relation obtained readily from Eq.\
\eqref{g2x2-ex}:
\begin{align}
 (\hat{g}_{2 \times 2} + i)
 \begin{bmatrix} -\cf \\ 1 \end{bmatrix}
 =
\begin{bmatrix} -\tf & 1 \end{bmatrix}
 (\hat{g}_{2 \times 2} + i) = 0.
\end{align}
This equation connects the diagonal element ($g$) and the off-diagonal
elements ($f$, $\widetilde{f}$) as
\begin{align}
 &g + i
 = \cf^{-1}f
 =- \tf^{-1}\widetilde{f},
 \label{gpi-f-rel}\\
 &g - i
 = \cf\widetilde{f}
 = -\tf f.
 \label{gmi-f-rel}
\end{align}
Adding \eqref{gpi-f-rel} and \eqref{gmi-f-rel}, we get
\begin{align}
 g
 &= \frac{1}{2}\left(\cf^{-1}f + \cf\widetilde{f}\right)
 \label{g-f-rel1}\\
 &= -\frac{1}{2}\left(\tf f + \tf^{-1}\widetilde{f}\right).
 \label{g-f-rel2}
\end{align}

It is worth noting that $|\cf|^2$ for $\epsilon=E$ satisfies
\begin{align}
 i\hbar v_F\hat{p}_z \partial_z |\cf|^2 = \Delta(z)(\cf^* - \cf) (1 - |\cf|^2).
 \label{abs-cf-eq}
\end{align}
Since $\cf(z \to +\infty) = \cf_2$ for $\hat{p}_z < 0$ and $|\cf_2|^2 = 1$
for $|E| < |\Delta_2|$, it follows from Eq.\ \eqref{abs-cf-eq} that
\begin{align}
 |\cf(\hat{p}, E, z)|^2 = 1 \ \
 (\hat{p}_z < 0,\ |E| < |\Delta_2|).
 \label{abs-cf-unity}
\end{align}
In the similar way, we can show that $\tf$ has the property
\begin{align}
 |\tf(\hat{p}, E, z)|^2 = 1 \ \
 (\hat{p}_z > 0,\ |E| < |\Delta_2|).
 \label{abs-tf-unity}
\end{align}

For the retarded Green's function $g(\hat{p},E+i0,z)$ at the low energies
$|E| < |\Delta_2|$, we obtain from Eqs.\ \eqref{abs-cf-unity} and
\eqref{g-f-rel1}
\begin{align}
 {\rm Im}[g] = {\rm Im}[\cf^* D] \ \ (\hat{p}_z < 0,\ |E| < |\Delta_2|)
 \label{Img-mpz}
\end{align}
and from Eqs.\ \eqref{abs-tf-unity} and \eqref{g-f-rel2}
\begin{align}
 {\rm Im}[g] = -{\rm Im}[\tf D] \ \ (\hat{p}_z > 0,\ |E| < |\Delta_2|),
 \label{Img-ppz}
\end{align}
where
\begin{align}
 D &= \frac{1}{2}[f(\hat{p},E+i0,z) - \widetilde{f}(\hat{p},E+i0,z)^*].
\end{align}
Since
\begin{align}
 \widetilde{f}(\hat{p}, E+i0, z)^*
 &= f(-\hat{p}, -E+i0, z) \notag \\
 &= f(\hat{p}, E-i0, z),
\end{align}
we can write $D$ in the form
\begin{align}
 D = \frac{1}{2}\left[f(\hat{p},E+i0,z) - f(\hat{p},E-i0,z)\right].
 \label{D0-ex}
\end{align}
Equations \eqref{Img-mpz}, \eqref{Img-ppz}, and \eqref{D0-ex} give the local
density of states, $n(\hat{p}, E, z) = {\rm Im}[g]$, at $|E| < |\Delta_2|$.

In the zero-energy limit, we have
\begin{align}
 &\lim_{E \to 0} D = f^{\rm OF}(\hat{p},\epsilon \to +i0,z),\\
 &\lim_{E \to 0} \cf = -i\,{\rm sgn}(\Delta_2) \ \ (\hat{p}_z < 0),\\
 &\lim_{E \to 0} \tf = i\,{\rm sgn}(\Delta_2) \ \ (\hat{p}_z > 0).
\end{align}
Substituting these expressions into Eqs.\ \eqref{Img-mpz} and
\eqref{Img-ppz} and considering that the density of states is positive
definite, we arrive at
\begin{align}
 n(\hat{p}, E \to 0, z)
 = \left|{\rm Re}[f^{\rm OF}(\hat{p},\epsilon \to +i0,z)]\right|.
\end{align}
This shows that the zero-energy density of states can be interpreted as a
manifestation of odd-frequency pairing.


\end{document}